\documentclass[submission]{eptcs}
\providecommand{\event}{Transformation Tool Contest 2013 (TTC'13)}

\usepackage{amsmath}
\usepackage{graphicx}
\usepackage{verbatim}
\usepackage{paralist}
\usepackage{listings}
\newcommand{\myemail}[0]{horn@uni-koblenz.de}

\title{The TTC 2013 Flowgraphs Case}
\author{Tassilo Horn
  \email{\myemail}
  \institute{University Koblenz-Landau, Institute for Software Technology, Germany}}
\def\titlerunning{The TTC 2013 Flowgraphs Case}
\def\authorrunning{Tassilo Horn}

\begin{document}

\maketitle

\begin{abstract}
  This case for the Transformation Tool Contest 2013 is about evaluating the
  scope and usability of transformation languages and tools for a set of four
  tasks requiring very different capabilities.  One task deals with typical
  model-to-model transformation problem, there's a model-to-text problem, there
  are two in-place transformation problems, and finally there's a task dealing
  with validation of models resulting from the transformations.

  The tasks build upon each other, but the transformation case project also
  provides all intermediate models, thus making it possible to skip tasks that
  are not suited for a particular tool, or for parallelizing the work among
  members of participating teams.
\end{abstract}

\section{Objective of the Case}
\label{sec:objective}

The objective of this case\footnote{This case's project on github:
  \url{https://github.com/tsdh/ttc-2013-flowgraphs-case}} is to evaluate the
flexibility of transformation tools, i.e., to evaluate their usefulness for
tasks requiring different capabilities.  Although different capabilities are
needed, all tasks are connected by their general topic: \emph{analysis and
  transformations in compiler construction}.

Task~1 deals with a typical model-to-model transformation problem.  Given an
abstract syntax graph of a Java program conforming to a very detailed
metamodel, the structure graph of the original program conforming to a much
simpler metamodel has to be generated.  Embedded in this task is a
model-to-text transformation where parts of the Java syntax graph have to be
serialized back to Java source code.

In task~2, the structure graph resulting from task~1 should be enhanced with
control flow information.  This is an in-place transformation task which is
suited for graph transformation tools but can also be tackled algorithmically.

Task~3 is also an in-place transformation.  Based on the control flow graph
resulting from task~2, data flow information has to be synthesized.  Again,
this task is suited to be tackled with graph transformations or
algorithmically.

The context of task~4 is a bit offside the strict transformation context.  A
simple validation tool and DSL should be developed to offload testing to Java
developers.

Because every task builds upon the results of previous tasks, the intermediate
models are also provided to allow participants to defer or skip tasks not
particulary suited for their tools, or to allow teams for developing solutions
in parallel.

\section{Detailed Task Description}
\label{sec:task-descr}

\paragraph{Task~1: Structure Graph.}
\label{sec:task1-structure-graph}

\begin{figure}[h!]
  \centering
  \includegraphics[width=.9\linewidth]{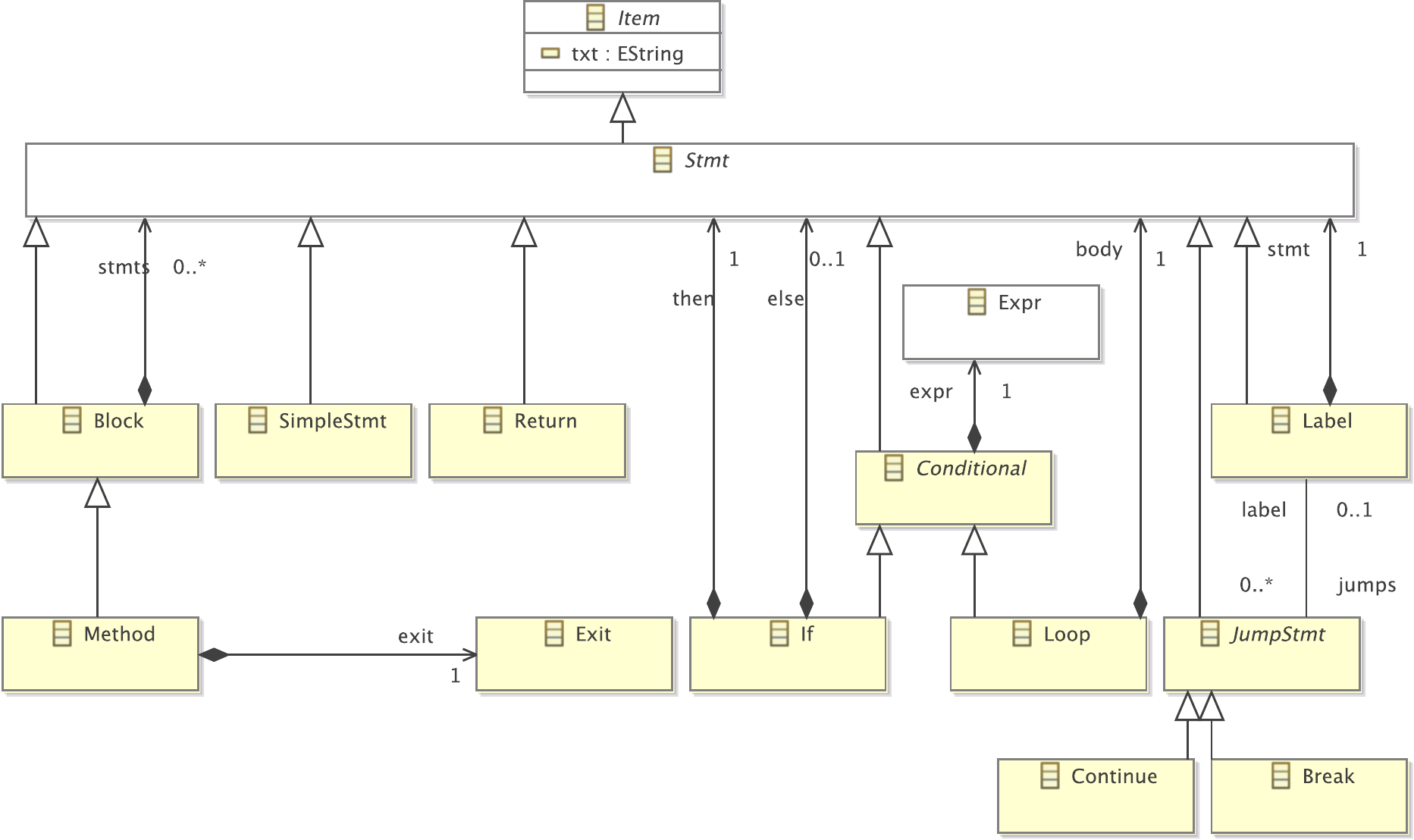}
  \caption{The target structure graph metamodel}
  \label{fig:structure-graph-mm}
\end{figure}

The first task requires writing a model-to-model transformation.  The source
models are abstract Java syntax graphs conforming to the JaMoPP metamodel
\cite{jamopp09}.  The JaMoPP metamodel covers the complete syntax of Java 7.
However, to restrict the size of the transformation, the elements actually
contained in the provided source models is limited.  They all contain one
compilation unit containing exactly one class with exactly one method.  The
method may have parameters.  In the method's body, there may be local variable
declarations, arithmetic expressions (only \verb|+|, \verb|-|, \verb|*|, and
\verb|/|), assignments, unary modification expressions (\verb|i++;| and
\verb|i--;|), \verb|return| statements, and blocks.  There may be
\verb|if|-statements and \verb|while|-loops with a boolean expression as
condition.  Statements may be labeled, and \verb|break|/\verb|continue| may be
used with or without target label.  The target metamodel of the transformation
is depicted in Figure~\ref{fig:structure-graph-mm}.

It is very similar to the original JaMoPP metamodel from a structural point of
view.  The major difference is that statements and expressions are represented
as one single object instead of being split up any further.  Another difference
is that every \verb|Method| has exactly one \verb|Exit|.  There is no
correspondence in Java, but it's a synthetic element added in favour of task~2.
No matter how a method is exited, the last object in a method's control flow
graph is this method's \verb|Exit| object.

All metamodel classes extend the abstract \verb|Item| class, even the class
\verb|Expr| although not visible in Figure~\ref{fig:structure-graph-mm}.
\verb|Item| declares a \verb|txt| attribute.  The transformation has to set the
value of this attribute to the concrete Java syntax of the statement or
expression, that is, there is a model-to-text transformation embedded in this
model-to-model transformation.

With the exception of \verb|Break| and \verb|Continue| objects that might refer
to a target \verb|Label|, the structure graphs created by the transformation
are simple trees that reflect the containment hierarchy of the method.

\paragraph{Task~2: Control Flow Graph.}
\label{sec:task2-cf-graph}

\begin{figure}[h!]
  \centering
  \includegraphics[width=0.7\linewidth]{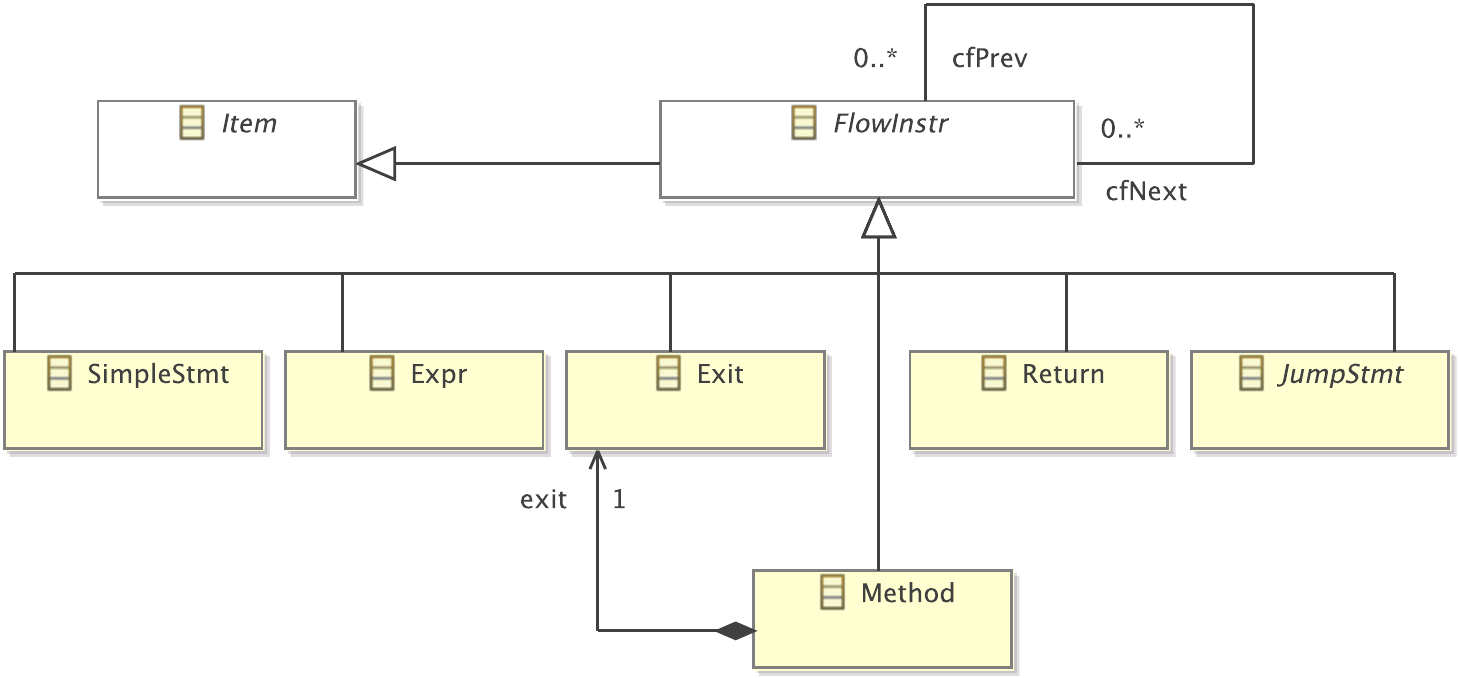}
  \caption{Metamodel classes related to control flow}
  \label{fig:control-flow-mm}
\end{figure}

This task deals with an in-place transformation problem.  The semantics of the
Java programming language should be integrated into the structure graphs
created by the previous transformation.  The task is to perform an
intra-procedural control flow analysis.  Any instruction should be connected to
the instructions that may follow it in the method's control flow.
Figure~\ref{fig:control-flow-mm} shows the relevant metamodel excerpt.

Simple statements, expressions, the synthetical exits, methods, \verb|return|,
and the jump statements \verb|break| and \verb|continue| extend
\verb|FlowInstr|.  Every flow instruction knows its immediate control flow
predecessors (\verb|cfPrev|) and successors (\verb|cfNext|).  It's those links
the transformation has to synthesize from the structure graph.

Blocks, labels, loops, and if-statements don't participate in the control flow.
Instead, when control flow reaches a block, the first flow instruction in the
block is the control flow successor of the previous flow instruction.  Since
blocks may be nested in other blocks, the \emph{first} flow instruction is
actually the first one reachable by a depth-first search.  These \emph{first}
semantics apply to the whole description of this task.

In case of a label, the first flow instruction in the labled statement is the
control flow successor.

In case of loops and if-statements, the successor is their test expression.
This expression has in turn two control flow successors.  If it is a test
expression of a loop, the successors are the first flow instruction in the
loop's body, and the first flow instruction following the loop.  If it is a
test expression of an if-statements, the first successor is the first flow
instruction in then-statement.  If there is an else-statement, its first flow
instruction is the other control flow successor.  Otherwise, the other
successor is the first flow instruction in the statement following the
if-statement.

The control flow successor of a \verb|Method| is its first flow instruction,
and \verb|Return| statements always have the method's \verb|Exit| as control
flow successor.

The most complex control flow rules apply to the \verb|Break| and
\verb|Continue| statements.  Without a target label, the control flow successor
of a \verb|Break| is the first flow instruction following the immediately
surrounding loop, and the successor of a \verb|Continue| is the test expression
of the immediately surrounding loop.  With a target label, the control flow
successor of a \verb|Break| is the first flow instruction following the labeled
statement, and the successor of a \verb|Continue| is the expression of the
surrounding labeled loop.

\paragraph{Task~3: Data Flow Graph.}
\label{sec:task3-df-graph}

\begin{figure}[h!]
  \centering
  \includegraphics[width=0.5\linewidth]{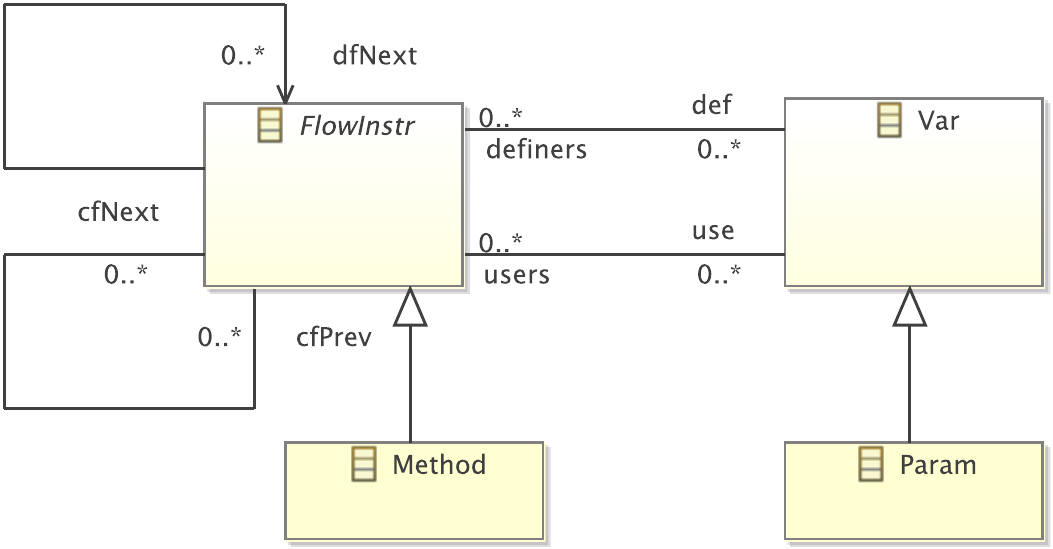}
  \caption{Metamodel classes related to data flow}
  \label{fig:data-flow-mm}
\end{figure}

In this task, an intra-procedural data flow analysis should be performed.  The
relevant metamodel excerpt is shown in Figure~\ref{fig:data-flow-mm}.  This can
be done based on the control flow graph, but one important piece of information
is missing from it: for every flow instruction, the sets of read and written
variables have to be known.  Therefore, this task is twofold:

\begin{compactenum}
\item The model-to-model transformation from task~1 has to be extended so that
  it also creates \verb|Var| objects for local variables and \verb|Param|
  objects for method parameters which are connected to flow instructions
  reading from and writing to them.
\item A transformation synthesizing data flow edges has to be written that
  takes the control flow graph resulting from applying task~2's transformation
  to the result of the extended java-to-structure-graph transformation.
\end{compactenum}

\subparagraph{Subtask~3.1.}
\label{sec:subtask-3.1}

For every local variable statement and every method parameter in the JaMoPP
model, the extended model-to-model transformation has to create a \verb|Var| or
a \verb|Param| object, respectively.  The \verb|txt| attribute should be set to
the variable's/parameter's name.  Furthermore, every flow instruction should be
connected to the variables it writes (the \verb|def| reference) and to the
variables it reads (the \verb|use| reference).

\subparagraph{Subtask~3.2.}
\label{sec:subtask-3.2}

The model resulting from applying the enhanced model-to-model transformation on
the JaMoPP syntax graphs followed by applying the control flow transformation
from task~2 to it is the source model for the data flow transformation to be
developed in this subtask.

It's sole purpose is to synthesize \verb|dfNext| links.  For every flow
instruction $n$, a \verb|dfNext| link has to be created from all nearest
control flow predecessors $m$ that define a variable which is used by $n$.
Formally:

\begin{align*}
  m \rightarrow_{dfNext} n  \iff {} & def(m) \cap use(n) \neq \emptyset\\
  ~\land {} & \exists~Path~m = n_0 \rightarrow_{cfNext} ... \rightarrow_{cfNext} n_k = n:\\
  & \left(def(m) \cap use(n)\right) \setminus \left(\bigcup_{0 < i < k}
    def(n_i)\right) \neq \emptyset
\end{align*}

That is, $n$ uses at least one variable defined by $m$, and there is a control
flow path from $m$ to $n$ in which at least one variable used by $n$ and
defined by $m$ is not redefined by intermediate flow instructions.

There are several ways to tackle this problem.  A simple one is to take the
definition literally, i.e., for every flow instruction search the nearest
control flow predecessors that define a variable used by instruction with
quadratic worst-case effort.  A more efficient and sophisticated algorithm is
described in the dragonbook \cite{Aho:CPTT}, chapter 9.1.  The models resulting
from this task which include control and data flow information are called
\emph{program dependence graphs} (PDG), and they play an important role in the
optimization phase in compilers \cite{Ferrante:1987:PDG:24039.24041}.

\paragraph{Task~4: Validation.}
\label{sec:task4-validation}

The fourth task is no strict transformation task.  Instead, the challenge is
validating the program dependence graphs resulting from task~3.  Concretely, it
should be checked if all \verb|cfNext| and \verb|dfNext| links are set
properly.

A simple tool that gets a result PDG as input and all control and data flow
links specified with a simple DSL should be provided.  The tool should print
all missing and all false links, i.e., all links defined in the textual
specification that don't occur in the model, and all links occuring in the
model that are not defined by the specification.

In the example Java programs provided in this case description project, every
statement and expressions occurs at most once in a method, e.g., there's is no
method with two \verb|i++;| statements.  Therefore, for all PDGs generated from
them, the \verb|txt| attribute can be used to uniquely identify any object.

An example specification is given in Listing~\ref{lst:example-validation}.
There's no restrictions on the actual syntax except that it should be easy to
write for a Java programmer.

\begin{lstlisting}[caption={An example validation DSL for result PDGs},label={lst:example-validation}]
                cfNext: "testMethod()"   --> "int a = 1;"
                cfNext: "int a = 1;"     --> "int b = 2;"
                ...

                dfNext: "int a = 1;"     --> "int c = a + b;"
                dfNext: "int b = 2;"     --> "int c = a + b;"
                ...
\end{lstlisting}

The requested tools's job is simple.  It has to load a PDG and to read a
specification such as depicted in Listing~\ref{lst:example-validation}.  For
any \verb|cfNext| or \verb|dfNext| link in the model, it has to check if it is
also defined in the specification.  If not, it has to print a \emph{false-link
  warning} message.  Reversely, it has to check if every link defined in the
specification also occurs in the model.  If not, it has to print a
\emph{missing-link warning}.

\section{Evaluation}
\label{sec:evaluation-criteria}

The evaluation of solutions has been done in two phases.  Before the workshop,
there was an open peer review where participants assessed the objective
criteria \emph{completeness}, \emph{correctness}, and \emph{efficiency}.
Furthermore, they assigned scores for the subjective criteria of the
transformation language's and tool's \emph{usefulness} and its \emph{ease of
  use}.  During the workshop, all attendants only assigned scores for the
subjective criteria.  The overall winner (the Epsilon solution) was then
determined by setting the open peer review scores off against the workshop
scores.

\bibliography{ttc-2013-flowgraphs-case}

\begin{thebibliography}{1}
\providecommand{\bibitemdeclare}[2]{}
\providecommand{\surnamestart}{}
\providecommand{\surnameend}{}
\providecommand{\urlprefix}{Available at }
\providecommand{\url}[1]{\texttt{#1}}
\providecommand{\href}[2]{\texttt{#2}}
\providecommand{\urlalt}[2]{\href{#1}{#2}}
\providecommand{\doi}[1]{doi:\urlalt{http://dx.doi.org/#1}{#1}}
\providecommand{\bibinfo}[2]{#2}

\bibitemdeclare{book}{Aho:CPTT}
\bibitem{Aho:CPTT}
\bibinfo{author}{Alfred~V. \surnamestart Aho\surnameend},
  \bibinfo{author}{Monica~S. \surnamestart Lam\surnameend},
  \bibinfo{author}{Ravi \surnamestart Sethi\surnameend} \&
  \bibinfo{author}{Jeffrey~D. \surnamestart Ullman\surnameend}
  (\bibinfo{year}{2006}): \emph{\bibinfo{title}{Compilers: Principles,
  Techniques, and Tools (2nd Edition)}}.
\newblock \bibinfo{publisher}{Addison-Wesley Longman Publishing Co., Inc.},
  \bibinfo{address}{Boston, MA, USA}.

\bibitemdeclare{article}{Ferrante:1987:PDG:24039.24041}
\bibitem{Ferrante:1987:PDG:24039.24041}
\bibinfo{author}{Jeanne \surnamestart Ferrante\surnameend},
  \bibinfo{author}{Karl~J. \surnamestart Ottenstein\surnameend} \&
  \bibinfo{author}{Joe~D. \surnamestart Warren\surnameend}
  (\bibinfo{year}{1987}): \emph{\bibinfo{title}{The program dependence graph
  and its use in optimization}}.
\newblock {\sl \bibinfo{journal}{ACM Trans. Program. Lang. Syst.}}
  \bibinfo{volume}{9}(\bibinfo{number}{3}), pp. \bibinfo{pages}{319--349},
  \doi{10.1145/24039.24041}.

\bibitemdeclare{techreport}{jamopp09}
\bibitem{jamopp09}
\bibinfo{author}{Florian \surnamestart Heidenreich\surnameend},
  \bibinfo{author}{Jendrik \surnamestart Johannes\surnameend},
  \bibinfo{author}{Mirko \surnamestart Seifert\surnameend} \&
  \bibinfo{author}{Christian \surnamestart Wende\surnameend}
  (\bibinfo{year}{2009}): \emph{\bibinfo{title}{{JaMoPP: The Java Model Parser
  and Printer}}}.
\newblock \bibinfo{type}{Technical Report} \bibinfo{number}{TUD-FI09-10},
  \bibinfo{institution}{Technische Universität Dresden, Fakult\"at
  Informatik}.
\newblock
  \bibinfo{note}{\url{ftp://ftp.inf.tu-dresden.de/pub/berichte/tud09-10.pdf}}.

\end{thebibliography}
\bibliographystyle{eptcs}
\end{document}